\documentclass[final,5p]{elsarticle}

\usepackage{amssymb}
\usepackage{subcaption}
\usepackage{amsthm}
\usepackage{multirow}
\usepackage[T2A]{fontenc}			
\usepackage[utf8]{inputenc}			
\usepackage{amsfonts,amssymb,amsthm,mathtools} 
\usepackage{amsmath}
\usepackage{booktabs}
\usepackage{color}

\usepackage{hyperref}
\hypersetup{
    colorlinks   = true,
    linkcolor    = black,
    citecolor    = blue,
    urlcolor     = blue
}

\usepackage[textwidth=1cm, textsize=scriptsize]{todonotes}

\journal{Acta Materialia}

\begin{document}

\begin{frontmatter}



\title{Adaptive fine-tuning of foundation models for crystal structure prediction: Discovery of high-pressure phases in the Ca--Fe--Ni system}

\author[inst1]{N.M. Chtchelkatchev}
\author[inst2]{M.V. Magnitskaya}
\author[inst3,inst4]{R.E. Ryltsev}

\affiliation[inst1]{organization={Joint Institute for Nuclear Research},
            addressline={Joliot-Curie St 6},
            city={Dubna, Moscow Region},
            postcode={141980},
            country={Russia}}

\affiliation[inst2]{organization={Vereshchagin Institute of High Pressure Physics, Russian Academy of Sciences},
            addressline={Kaluzhskoe sh. 14},
            city={Moscow (Troitsk)},
            postcode={108840},
            country={Russia}}

\affiliation[inst3]{organization={Vatolin Institute of Metallurgy of the Ural Branch of the Russian Academy of Sciences},
            addressline={Amundsen str. 101},
            city={Ekaterinburg},
            postcode={620016},
            country={Russia}}

\affiliation[inst4]{organization={Ural Federal University},
            addressline={Lenin Ave, 51},
            city={Ekaterinburg},
            postcode={620002},
            country={Russia}}

\cortext[cor1]{Corresponding address: shchelkachev.nm@phystech.edu}

\begin{abstract}
The prediction of crystal structures remains a central challenge in chemistry and materials science, underpinning the discovery of materials with targeted properties. Although evolutionary algorithms efficiently explore complex energy landscapes, their reliance on \textit{ab initio} calculations for structural relaxation and ranking leads to prohibitive computational costs. Machine learning interatomic potentials (MLIPs) can substantially accelerate crystal structure prediction (CSP), yet their practical integration into CSP workflows remains limited by the need for large, carefully selected training datasets. In particular, an outstanding challenge is how to identify, from the vast number of candidate structures generated during global searches, a compact and informative subset that warrants expensive density functional theory (DFT) calculations for model refinement. Recent advances in foundation models, pretrained on large-scale \textit{ab initio} databases spanning millions of atomic structures across the periodic table, enable accurate MLIPs to be constructed with substantially reduced computational effort via transfer learning. Here we develop a self-consistent foundation-model-assisted CSP workflow that couples evolutionary search with an adaptive data-selection and fine-tuning strategy. The proposed algorithm simultaneously performs rapid structure exploration and constructs a system-specific ML potential by iteratively selecting representative and physically relevant configurations for DFT labeling, thereby minimizing redundant computations while maintaining high accuracy. Applied to the chemically complex Ca--Fe--Ni ternary system, the approach accurately reproduces the known low-pressure convex hull, validating the methodology, and enables efficient exploration of high-pressure phases. Using this framework, we predict a previously unknown compound, Ca$_6$FeNi, which becomes thermodynamically stable above 100~GPa and has not yet been reported experimentally. These results demonstrate that foundation-model-based, data-efficient workflows can both accelerate crystal structure prediction and enable the discovery of genuinely new materials in complex multicomponent systems.
\end{abstract}

\begin{graphicalabstract}
\includegraphics{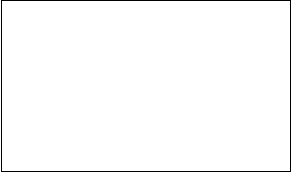}
\end{graphicalabstract}

\begin{highlights}
\item Developed a self-consistent workflow coupling evolutionary search with foundation models
\item  Introduced data-efficient selection of representative structures for DFT fine-tuning
\item  Reproduced the known low-pressure convex hull of the Ca--Fe-–Ni system
\item Predicted two binary compounds Ca$_3$Fe$_2$ and Ca$_8$Fe$_5$ above 50~GPa and one ternary compound Ca$_6$FeNi above 100~GPa.
\end{highlights}


\end{frontmatter}


\section{INTRODUCTION}
\label{sec:intro}


Crystal structure prediction (CSP) is a cornerstone of modern materials science, providing a systematic route for discovering stable and metastable phases and for understanding structure--property relationships across wide ranges of composition, pressure, and temperature \cite{Oganov2019,Woodley2008}. In particular, CSP plays a crucial role in regimes where experimental characterization is challenging or infeasible, such as at high pressures or for complex multicomponent systems, where kinetic barriers and metastability strongly affect phase formation \cite{Zurek2015,Wang2014}.


Over the past two decades, global optimization techniques, including evolutionary algorithms and related methods, have demonstrated remarkable success in identifying low-energy crystal structures directly from chemical composition \cite{Oganov2006,Wang2012,Lyakhov2013}. However, such approaches are traditionally coupled with \emph{ab initio} density functional theory (DFT) calculations for structural relaxation and energetic ranking, which leads to prohibitively high computational costs when exploring large compositional spaces, multicomponent systems, or high-pressure phase diagrams \cite{Kruglov2023}. These limitations become especially severe when high accuracy is required, for instance for constructing reliable convex hulls or resolving small enthalpy differences between competing phases.

Machine-learning interatomic potentials (MLIPs) have emerged as a powerful solution to this computational bottleneck, offering near-DFT accuracy at a fraction of the computational cost \cite{Behler2016,Unke2021}. Their integration into CSP workflows enables large-scale structure screening, accelerated evolutionary searches, and efficient prescreening of candidate phases \cite{Zhang2019,Smith2018}. Nevertheless, the construction of robust MLIPs typically relies on extensive system-specific training datasets and iterative active learning schemes, which themselves require substantial \emph{ab initio} labeling~\cite{Kellner2024,Bigi2024}. This creates a fundamental challenge for previously unexplored materials or extreme thermodynamic conditions, where representative training data are scarce or entirely unavailable.

Recent progress in so-called foundation models (FMs) for atomistic simulations has opened a new paradigm for CSP and materials discovery \textcolor{red}{\cite{Batatia2022,Batatia2023}}. These models, pretrained on large-scale and chemically diverse \emph{ab initio} datasets encompassing millions of atomic configurations, provide transferable representations of interatomic interactions across wide regions of chemical and structural space. FMs based on equivariant graph neural networks have demonstrated strong generalization capabilities and promising accuracy for a broad range of materials classes \textcolor{red}{\cite{Thomas2023,Batatia2024}}. Importantly, such models can be efficiently adapted to specific target systems through transfer learning or fine-tuning, dramatically reducing the amount of system-specific DFT data required.


However, integrating FMs into CSP remains nontrivial. Evolutionary searches typically generate thousands of candidate structures, many of which are redundant, high-energy, or physically irrelevant. Directly labeling all such configurations with DFT is computationally infeasible, while indiscriminate selection often degrades the quality and transferability of the resulting potential. Consequently, a central algorithmic challenge is to identify a compact yet informative subset of structures that maximally improves the potential during fine-tuning while minimizing expensive \textit{ab initio} calculations. Designing such data-efficient selection strategies is essential for making FM-assisted CSP both accurate and scalable, yet remains largely unexplored. This problem become even more challenging under extreme pressures, as most large-scale training datasets are dominated by near-ambient conditions and so stability of FM at such conditions is an open issue.

In this work, we develop a self-consistent and computationally efficient CSP workflow that combines evolutionary structure search with state-of-the-art FMs and transfer learning, and apply it to the Ca--Fe--Ni ternary system over a wide pressure range. Iron and nickel are the principal constituents of the Earth's core, governing its mechanical, thermal, and magnetic properties \cite{Dziewonski1981,McDonough1995}. The well-known density deficit of the core compared to the density of iron and iron-nickel alloys suggests the presence of lighter element impurities in the core. Hypotheses about the composition of the Earth's core are based on indirect data on the abundance of chemical elements in the lithosphere and in space. Traditionally, light non-metallic elements such as H, C, O, Si, S, and some other, less common, non-metals are considered as impurities in the core \cite{Dziewonski1981,McDonough1995,Litasov2016}. Despite numerous studies, the question of which light elements are present in the Earth's outer or inner core remains controversial. To resolve this issue, it is necessary, in particular, to understand which iron-based chemical compounds are possible under conditions characteristic of the Earth's interior, and how their stability, structure, and properties change under the influence of high pressure.

Sometimes the alkali metal potassium, which does not chemically bind to iron group metals at normal pressure, is also considered as a candidate trace element present in the earth's core (see e.g. \cite{Wood2006,Watanabe2014}). This is because the decay of the $^{40}$K isotope is considered one of the sources of the earth's radiogenic heat (along with the decay of U and Th). Possibility of K alloying with 3d-metals was proved experimentally in \cite{Parker1996}, where the ordered compound between K and Ni was synthesized in diamond anvils at $P$=37~GPa.

Our work is motivated by the assumption, first put forward in \cite{Tsvyash1998,Tsvyash2002}, that alkaline earth calcium, the fifth most abundant element in the lithosphere, may also be present in the core. This is because calcium's most abundant isotope, $^{40}$Ca, is a product of the spontaneous $\ensuremath{\beta}$-decay of $^{40}$K, and calcium is even more widespread in the lithosphere than potassium. Furthermore, recent publications (e.g., \cite{Yan2023,DongGX2024}) note that calcium is likely more abundant in the universe than traditionally believed, since Ca was formed in large quantities in the first stars through the interaction of protons and the isotope $^{19}$F, and the frequency of these reactions was previously greatly underestimated.

Being, like K, a lithophile element, Ca does not form ordered compounds or solid solutions with Fe at ambient pressure. However, it forms chemical compounds with Ni, a 3d-element also related to the Earth's core. Moreover, an ordered compound of calcium with another iron group metal cobalt --- CaCo$_2$ with the C15 structure --- was synthesized for the first time under high pressure of 8~GPa \cite{Tsvyash1998}. The formation of CaCo2 is mainly a consequence of the pressure-driven changes in calcium's properties. From the viewpoint of band theory, the pre-transition element Ca undergoes the electron charge transfer from the 4s-band to the initially empty 3d-band. More precisely, there is a continuous change in the symmetry of the wave function of valence electrons due to pressure-induced increase in the contribution of states with higher orbital momenta, since the $l > 0$ states are less compressible than the spherical s-states (see, for example, review \cite{Maksimov2005}). As a result, such s-block elements as K and Ca become effectively transition metals and their affinity for Fe increases upon compression.

Intuitively, the formation of compounds of s-metals with Fe (and other iron group metals) is hindered by the fact that their atomic radii greatly exceed that of Fe. But since their compressibility is approximately an order of magnitude higher than that of Fe, the difference in atomic sizes is sharply reduced upon compression. Thus, for a binary compound to form, the difference in atomic radii of its constituents should not be too large. Another key parameter that determines the alloying of two elements is the difference in their electronegativities (work functions), which must be sufficiently large. This rule was first formulated in numbers as the semi-empirical Miedema criterion \cite{Miedema1975,Pettifor1995} and was subsequently significantly reformulated and refined \cite{Rahm2019,RahmZeng2019,Allahyari2020,Rahm2021,Dong2022}. A recent example is the work \cite{Kostenko2022}, where the authors propose an original method for predicting formation of many binary compounds for the entire periodic table. In \cite{Rahm2021,Dong2022}, the pressure-induced variations in electronegativity were calculated for many elements over the megabar pressure range. Qualitatively, these data imply that Ca should chemically bind with Ni, Co, and Fe at normal, moderate, and elevated pressure, respectively. The reliability of this estimate can be verified using precise calculations of the Ca--Fe--Ni system over a wide pressure range.

Here, we apply our FM-assisted CSP approach to accurately and reliably determine stable phases on a convex hull in the Ca--Fe--Ni system at various pressures. Somewhat unexpectedly, difficulties arise in accurately determining unary phases of elemental Ca. Сrystalline calcium has a very rich and complex phase diagram, the underlying cause of which is the pressure-induced changes in calcium's electron spectrum due to the s-d transition. In particular, recent simulations predict the formation of a weak electride state in compressed Ca \cite{Dong2017,Novoselov2020,Modak2023,Racioppi2025}. The spectrum of fcc-Ca is characterized by topological Dirac nodal lines near the Fermi level, and at moderate pressures, Ca becomes a nodal line semimetal \cite{Hirayama2017}. Many experimental and theoretical studies disagree on such issues as a pressure range of metallicity loss in fcc-Са (see e.g.\cite{Maksimov2005,Brazhkin2006,Magnitskaya2014}), the stability regions of various phases, such as bcc and sc at $P < 50$~GPa \cite{Modak2023,Oganov2010}, the transition to an incommensurate host-guest structure at $P > 100$~GPa \cite{Arapan2008,Oganov2010}, etc.

In this context, the Ca--Fe--Ni ternary system represents a particularly compelling and challenging test case. From a materials science perspective, the Ca--Fe--Ni system also exemplifies a complex multicomponent landscape characterized by competing bonding motifs, large atomic size mismatch, and pressure-driven changes in chemical behavior, making it an ideal benchmark for advanced CSP methodologies.

Geochemically, our calculated results show that the lithophile element Ca gradually becomes siderophile with increasing pressure.

Our results highlight both the capabilities and the limitations of FM-assisted CSP under extreme conditions and provide new insights into the high-pressure phase stability of Ca--Fe--Ni.

The results obtained for the Ca--Fe--Ni system improve our understanding of the poorly studied high-pressure chemistry of Ca and support the suggestion that Ca may be present in the Earth's core as a trace element.


\section{Methods}

\subsection{Accuracy requirements for foundation models under high pressure}

Evolutionary algorithms is an efficient approach for exploring stable crystal structures of materials over a wide pressure range. In this work, we study ternary Ca--Fe--Ni system in the pressure range (0--100~GPa) and employ the USPEX algorithm, which has proven highly effective for global structure searches across diverse chemical systems \cite{Oganov2006,Lyakhov2013}.

To accelerate structure relaxation and energy evaluation during the search, one can use MLIPs to describe interatomic interactions. Here we consider foundation models (FMs) based on graph neural networks, which provide near-\textit{ab initio} accuracy at a fraction of the computational cost \cite{Behler2016,Unke2021,Batatia2023}. Such models are pretrained on large and chemically diverse datasets (e.g., Materials Project, OMAT24, Alexandria), comprising millions of crystalline and molecular configurations and covering most of the periodic table \textcolor{red}{\cite{Jain2013,Alexandria2024,OMAT2024}}.

The key question is if these universal models provide sufficient accuracy for evolutionary searches and have been successfully applied to various materials systems. In that context ''sufficient'' accuracy means that true ground state structures are among the list of structures in a relatively narrow vicinity of the convex hull (say 0.1 eV/atom) generated by EA with using FM as energy calculator. In this case we can perform DFT refinement of reasonable number of structures to find true convex hull.

Thus, we expect that, at near-ambient pressures, FMs meet the above requirement, at least for certain classes of materials. Indeed, according to Matbench Discovery --  framework for evaluation FMs -- the accuracy of state-of-the-art models reach 0.02 eV/atom for thermodynamic stability and formation energy of inorganic materials throughout the periodic table~\cite{Riebesell2025MatMachInt}.

\begin{figure}[tb]
    \centering
    \includegraphics[width=0.45\textwidth]{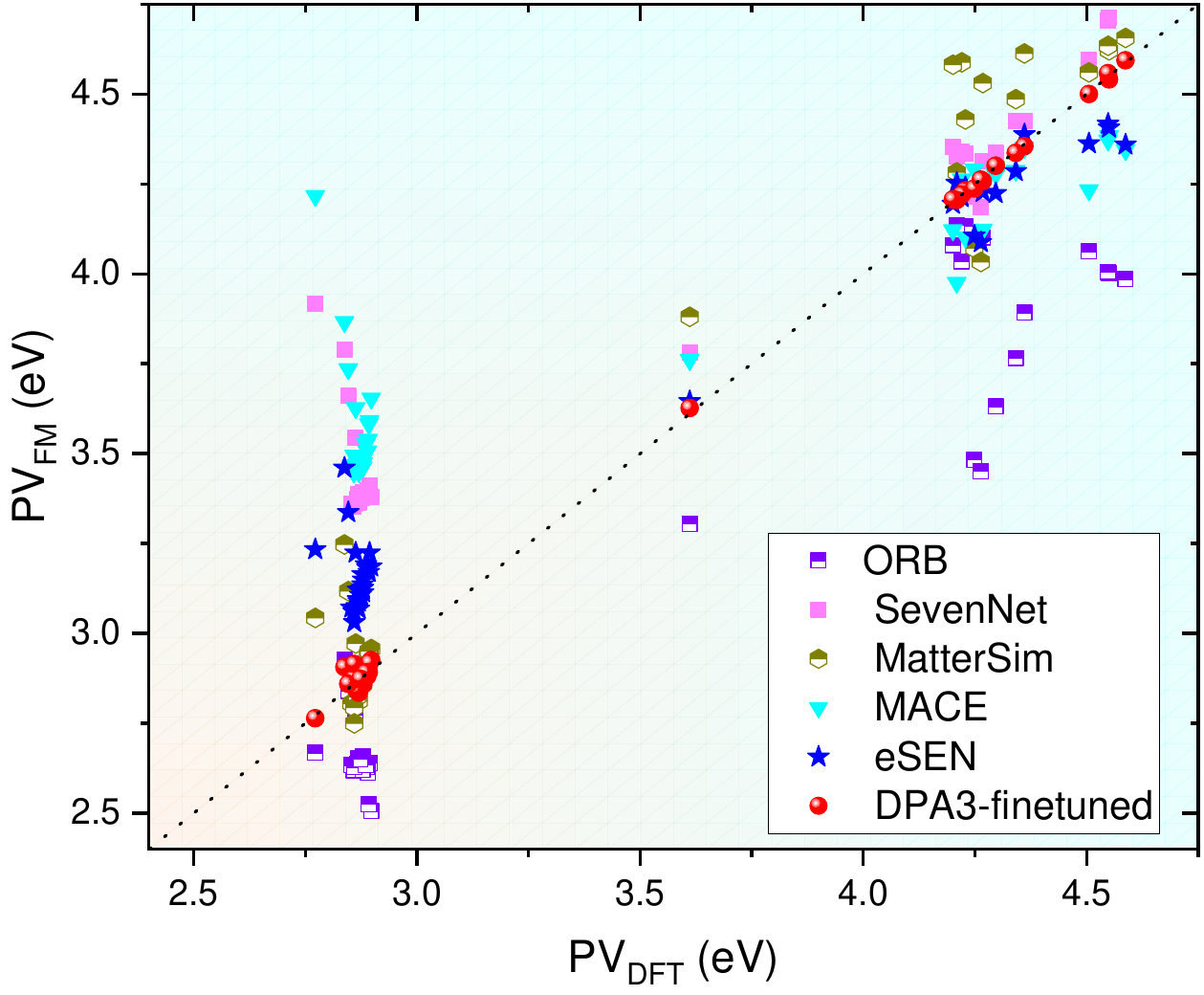}
    \caption{Parity plots FM vs DFT for $PV$ (per atom) calculated for the structures from extended convex hull in Ca-Fe-Ni  system at $P=50$~GPa. We see that all tested FM fail at elevated pressures whereas system specific finetuned version of DPA3 provide enough accuracy.}
    \label{fig:MLP_vs_vasp_10GPA}
\end{figure}

At elevated pressures, however, we expect that FMs may fail in evaluating stability of materials, because such high-pressures structures belong to extrapolation domain in comparison to training data. This is especially true for the system demonstrating substantial change of electronic structure under pressure (like Fe-Ni-Ca under investigation).

At elevated pressures, phase stability is governed by the enthalpy
\begin{equation}
    H = E + PV,
\end{equation}
rather than the internal energy alone. Even moderate errors in predicted energies or volumes can therefore lead to substantial errors in $PV$ and, consequently, incorrect phase ranking. A simple order-of-magnitude estimate shows that for a typical atomic volume of $V \sim 20$--30~\AA$^3$, the $PV$ contribution reaches $\sim0.1$~eV/atom already at pressures of $\sim$1~GPa. Thus, sub-0.1~eV accuracy per atom is required for reliable high-pressure structure selection.

Using pretrained FMs ``as is'' is acceptable only if this accuracy is maintained. Otherwise, extensive \textit{ab initio} refinement becomes unavoidable. For the Ca--Fe--Ni system, post-processing $\sim$1000 candidate structures after each USPEX run would require on the order of $5\times10^4$ DFT calculations, rendering brute-force validation impractical. Training a system-specific MLIP from scratch would require an additional $10^3$--$10^4$ labeled configurations. These considerations motivate the development of an accurate, fine-tuned model tailored to the target chemistry.

Figure~\ref{fig:MLP_vs_vasp_10GPA} and Table~\ref{tab:FM} compare the accuracy of several publicly available FMs at $P=50$~GPa using 44 representative structures from the extended convex hull. Enthalpy contributions ($PV$) predicted by the models are benchmarked against DFT calculations performed with VASP. We see that all the tested universal models exhibit maximal errors approaching or exceeding 1~eV/atom and MAE more than 0.1 eV/atom, which is insufficient for reliable phase ranking under compression. Among the tested models, MatterSim shows the best transferability; nevertheless, its accuracy remains inadequate for quantitative CSP. Thus, we conclude that to achieve the required precision, we have to construct a system-specific potential by fine-tuning appropriate FM. Below we will describe the pipeline for this procedure for DPA3 FM \cite{Zhang2018,DeepMD2020}; the results of the test for this model is shown in Figure~\ref{fig:MLP_vs_vasp_10GPA}. We see that finetuned model provides maximum $PV$ error below 0.07~eV/atom and MAE of 0.012~eV/atom, enabling robust high-pressure structure selection and so significantly reducing the number of required DFT evaluations.

\begin{table}[tbh]
    \centering
    \begin{tabular}{l | l | c}

        Name, Ref. & Model & MAE, eV\\
        \hline
        \hline
       eSEN  & eSEN-30M-OAM & 0.204 \\
       ORB   & ORB v3 & 0.328\\
       SevenNet & SevenNet-MF-ompa & 0.399\\
       MACE  & MACE-MPA-0 &  0.481\\
       MatterSim & MatterSim v1 5M & 0.104 \\
       DPA3 & DPA3-finetuned & 0.012 \\
        \hline
    \end{tabular}
    \caption{The list of tested FM with MAE for the $PV$ values calculated with respect to DFT. Current rating of these models can be found in Matbench Discovery platform~\cite{Matbench}. }
    \label{tab:FM}
\end{table}

\subsection{Structure generation, relaxation protocol,\\ and configuration-space coverage}

The construction of a reliable training database for fine-tuning a system-specific foundation model (FM) requires representative sampling of the relevant configuration space, including diverse stoichiometries, atomic environments, and compression regimes spanning $0$–$500$~GPa.

\begin{figure*}[th]
    \centering
    \includegraphics[width=0.8\textwidth]{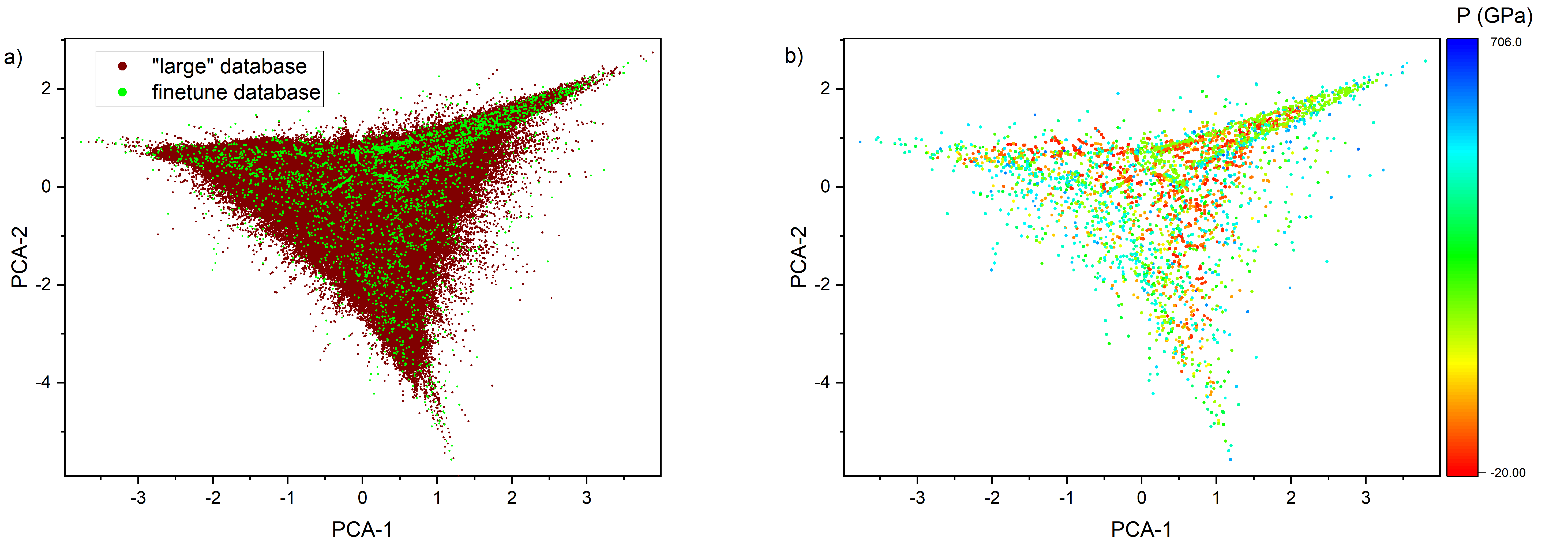}
    \\
    \includegraphics[width=0.8\textwidth]{PCA_concentrations.png}
    \caption{Two-dimensional PCA map of structures generated for Fe-Ni-Ca system by USPEX algorithm. (a) Structures predicted with MatterSim (red dots) as well as structures selected for further DFT labeling (green dots); (b) Structures selected for labeling colored by means of corresponding pressures; (c)-(e) Concentration distribution of elements for selected structures.}
    \label{fig:PCA}
\end{figure*}

\paragraph{Two-stage relaxation protocol}

Evolutionary algorithms such as USPEX frequently generate highly distorted trial structures, including configurations with unrealistically short interatomic distances and extremely large forces (hundreds of eV/\AA). Direct relaxation of such structures with high-accuracy FMs may lead to numerical instabilities or inefficient convergence.

To ensure robust and physically consistent relaxation, we employed a two-stage protocol using two complementary pretrained FMs.

In the first stage, structures were relaxed using the MACE FM, which has demonstrated improved robustness under extreme compression conditions due to its local message-passing architecture and stable force extrapolation behavior at short interatomic distances. This makes it well suited for preconditioning highly distorted configurations. Initial relaxation with MACE was performed with FIRE algorithm (ASE implementation) until the maximum atomic force decreased below $1$~eV/\AA.

In the second stage, relaxation was continued using the MatterSim FM, which provides superior energetic accuracy for near-equilibrium configurations in the Ca–Fe–Ni system (see Fig.~\ref{fig:MLP_vs_vasp_10GPA}). MatterSim relaxation proceeded in two substages. The first one was performed with FIRE algorithm until the maximum force $f_{\rm max}$ was below $0.1$~eV/\AA, and the second one with BFGS algorithm until $f_{\rm max} < 0.01$~eV/\AA.

This staged approach separates two numerical tasks:
(i) stabilization of highly nonequilibrium structures, and
(ii) accurate final relaxation near local minima.
The protocol significantly improves convergence reliability while maintaining computational efficiency for thousands of candidate structures.

All structure relaxations were performed using the Atomic Simulation Environment (ASE). A lightweight Python interface was developed to connect USPEX to ASE-based FM calculators, enabling automated structure optimization within the evolutionary search loop. Technical implementation details are provided in the Supporting Information.

Using the above relaxation protocol we performed evolutionary USPEX-based crystal structure search at $P = 0$, $50$, and $100$~GPa. For each pressure, 20 generations of 256 structures were explored, with unit cells containing 2–24 atoms. Initial seeds included known phases of elemental Ca, Fe, Ni, and binary compounds from the Materials Project database. Highly anisotropic cells (one lattice vector exceeding the others by more than a factor of three) were discarded to avoid quasi-one-dimensional artifacts.

Since we aimed to perform both CSP and fine-tuning system-specific FM, we used augmentation and local enrichment of structures obtained by USPEX. Thus, USPEX dataset was augmented via structures found for Ca--Fe--Ni in Materials Project, OQMD ans AFLOW databases. For each database structure incorporated into the training set, at least five perturbed configurations were generated by applying small random atomic displacements (amplitude $0.1$~\AA) and random lattice strains up to $5\%$. This augmentation serves two purposes: (i) it stabilizes fine-tuning by sampling the local curvature of the potential energy surface around known minima, (ii) it improves force and virial training by introducing controlled off-equilibrium configurations. The resulting dataset therefore contains both equilibrium structures and systematically perturbed configurations, enhancing transferability across pressure regimes.

To quantify the coverage of the configuration space and guide DFT data selection, we employed configuration-space analysis. For this purpose, we applied ASAP library~\cite{Cheng2020AccChemRes} implementation of Principal Component Analysis (PCA) in the structural space of Smooth Overlap of Atomic Positions (SOAP) descriptors. SOAP provides a rotationally and translationally invariant representation of local atomic environments, enabling quantitative comparison between structurally distinct configurations. For each structure, atomic SOAP descriptors were computed using a cutoff radius $r_{\mathrm{cut}} = 4.2$~\AA, $n=4$ radial basis functions, maximum angular momentum $l=3$, and Gaussian width $\sigma = 0.52$~\AA. A structure-level descriptor was obtained by averaging atomic descriptors over all atoms in the unit cell. The resulting high-dimensional descriptor space was projected onto its first two principal components to visualize structural diversity and detect clustering.

\begin{table*}[ht]
\centering
\caption{Accuracy of the fine-tuned DeepMD–DPA3 foundation model at different pressures.
The test set consists of structures generated by USPEX and verified by DFT calculations for convex-hull construction.
$N_{dE<50\,\mathrm{meV}}$ denotes the number of structures predicted by DPA-3 whose energy above the convex hull is below 50~meV/atom and which were therefore selected for DFT refinement.
$N_{dE<\mathrm{RMSE}}$ denotes the number of structures whose energy above the convex hull is smaller than the RMSE of the model at the corresponding pressure.}
\label{tab:dpa3_ac}

\begin{tabular}{lcccccccc}
\toprule
Pressure & \multicolumn{2}{c}{Energy (meV/atom)}
         & \multicolumn{2}{c}{Force (meV/\AA)}
         & \multicolumn{2}{c}{Virial (meV/atom)}
         & \multicolumn{2}{c}{Convex-hull structures} \\
(GPa)    & MAE & RMSE & MAE & RMSE & MAE & RMSE & $N_{dE<50}$ & $N_{dE<\mathrm{RMSE}}$ \\
\midrule
0   & 33.7 & 41.0 & 25.3 & 46.1 & 16.5 & 30.8 & 432 & 339 \\
50  & 16.9 & 21.5 & 16.7 & 27.2 & 8.5  & 14.5 & 329 & 79  \\
100 & 15.3 & 20.3 & 47.3 & 75.8 & 21.0 & 42.4 & 378 & 116 \\
200 & 7.3  & 9.6  & 24.7 & 36.1 & 10.0 & 18.2 & 148 & 18  \\
\midrule
Total & 11.3 & 47.7 & 79.6 & 119.0 & 11.9 & 78.9 & 1287 & 552 \\
\bottomrule
\end{tabular}
\end{table*}
\begin{figure*}[th]
    \centering
    \includegraphics[width=0.95\textwidth]{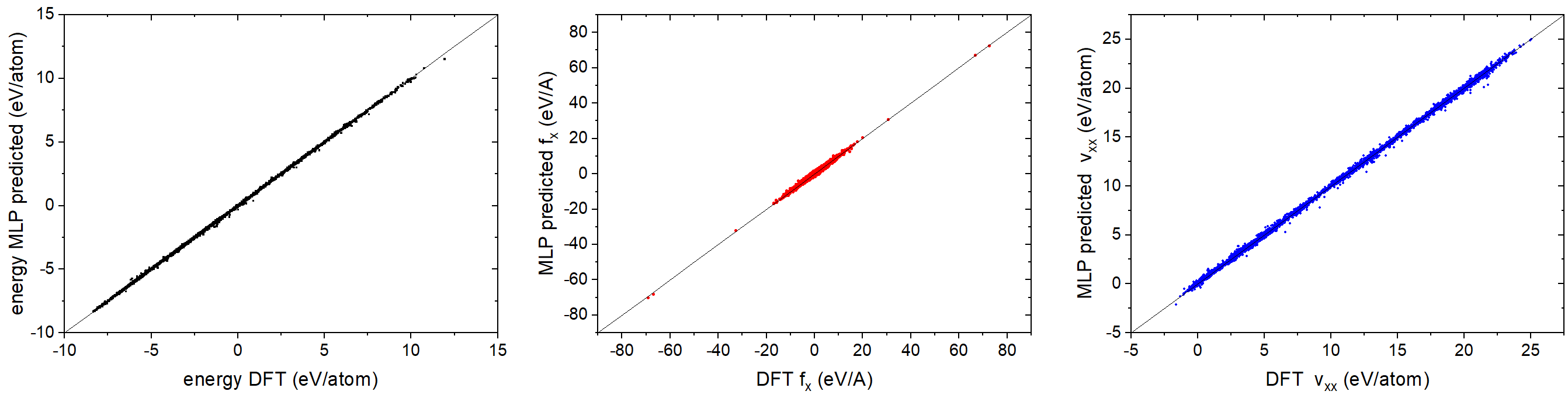} \\
    \caption{ Parity plots illustrating the accuracy of the fine-tuned DeepMD–DPA3 foundation model.
DFT reference values of energy, forces, and virial are shown on the horizontal axis, while the corresponding FM predictions are shown on the vertical axis. Deviation of the data points from the diagonal line $y = x$ quantifies the prediction error. The plots demonstrate that the model reproduces all three target quantities over a wide range of values with moderate deviations, confirming its suitability for atomistic simulations under the investigated conditions.}
    \label{fig:MLP_acc}
\end{figure*}
\begin{figure}[tbh]
    \centering
    \includegraphics[width=0.7\columnwidth]{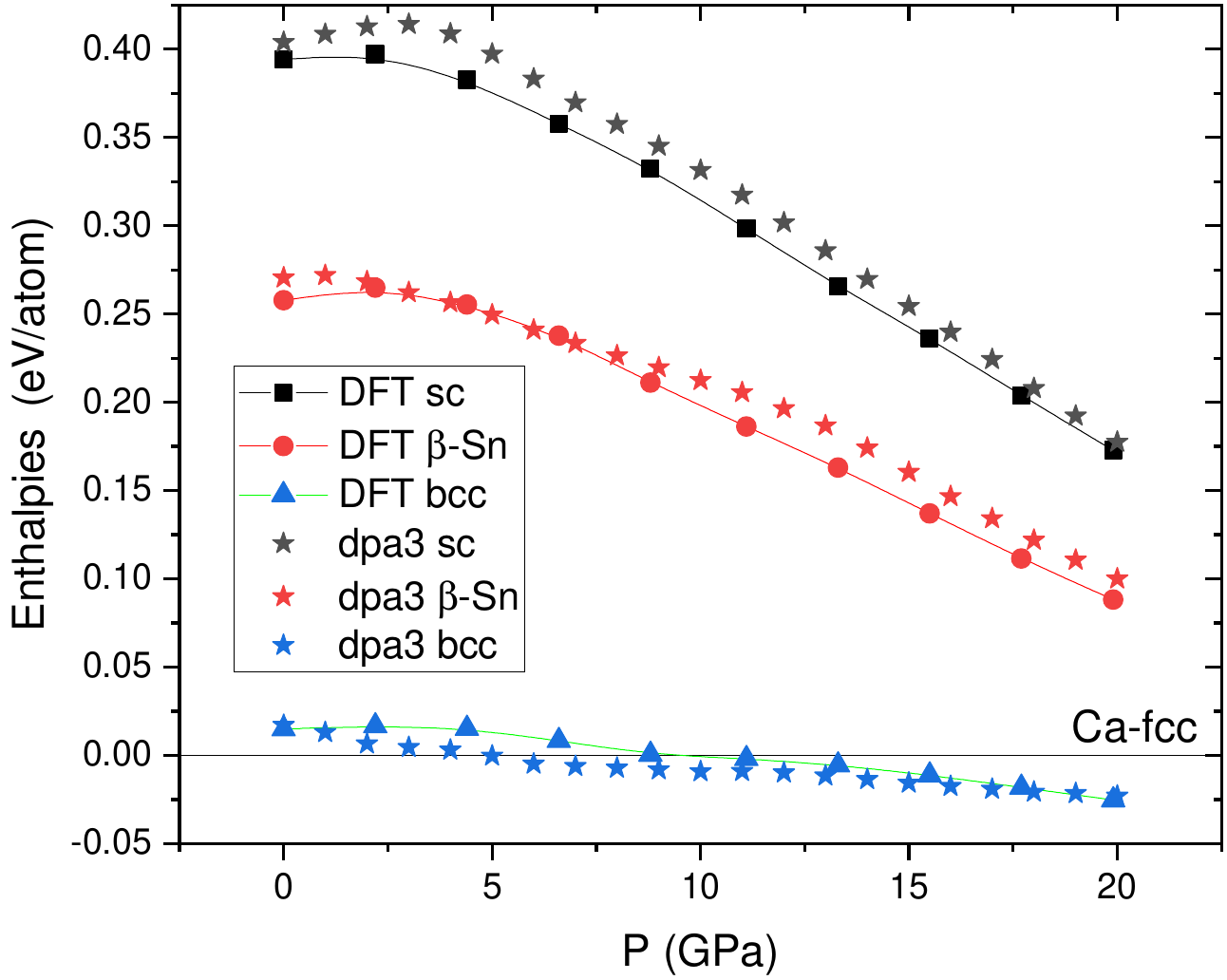} \\
    \caption{ Pressure dependence of relative enthalpies of the bcc, simple cubic, and $I4_1/amd$ ($\beta$-Sn prototype) structures with respect to the fcc phase, obtained from DFT and the DPA-3 foundation model.}
    \label{fig:Ca_P0-20}
\end{figure}

In Fig.~\ref{fig:PCA} we show the result of PCA analysis for the structures generated by USPEX algorithm.
The PCA projection demonstrates broad and approximately uniform coverage of the accessible configuration space. Clustering effects, which are seen for selected configurations reflect effects of dataset augmentation described above.

The pressure distribution shown in Fig.~\ref{fig:PCA}(b) exhibits a non-uniform but physically reasonable behavior, without pronounced clustering within a narrow region of the configuration space. This supports the representativeness of the selected dataset across the investigated pressure range. Figures~\ref{fig:PCA}(c)–(e) present the spatial distribution of Ca, Fe, and Ni concentrations in the selected structures, projected onto the principal component space. Notably, regions corresponding to the vertices of the compositional triangle in the PCA representation do not coincide with pure elemental phases, as might be expected from conventional ternary phase-diagram representations. Instead, these regions are characterized by minimal concentrations of the respective element. This behavior reflects the nontrivial chemical complexity of the ternary Ca–Fe–Ni system and the presence of strong interatomic interactions across broad compositional ranges. The observed distribution is consistent with the formation of stable intermetallic compounds and complex solid solutions, where even low concentrations of a given element can substantially influence the electronic structure and phase stability.

\subsection{Density Functional Theory Calculations}
\begin{figure*}[th]
    \centering
    \includegraphics[width=0.8\textwidth]{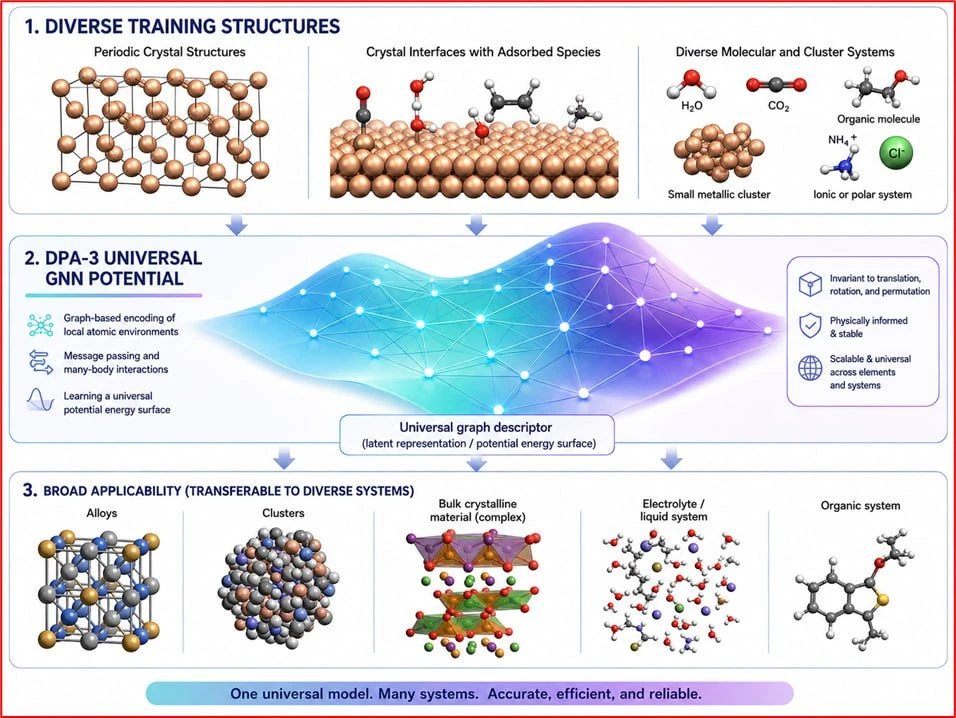} \\
    \caption{ The architecture of the Universal Graph-basedNeural Network Potential (UGLMP) DEEPMD-dpa3 is shown. The descriptor, implemented as a graph neural network, was pretrained on combined datasets containing diverse atomic configurations (crystals, molecules, clusters, etc.). The choice of the most suitable transformation network (the “fitting net”), which maps the descriptor output to the system’s potential energy, depends on the specifics of the task. In this work we used the “MP” configuration --- a transformation network pretrained on structures from the Materials Project database --- because our calculations conform to the methodological standards of that database.
}
    \label{fig:architecture}
\end{figure*}

\begin{figure*}[th]
    \centering
    \includegraphics[width=0.8\textwidth]{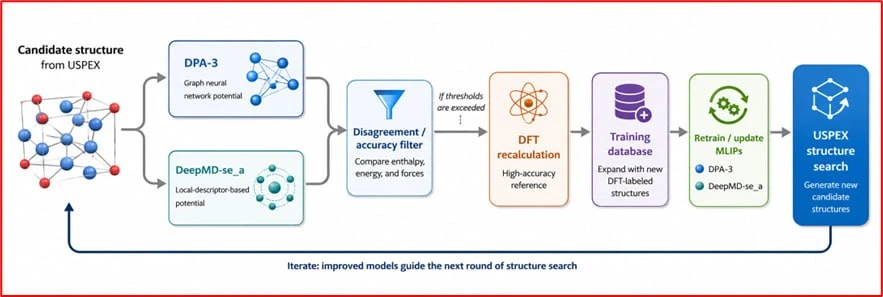} \\
    \caption{ Each structure predicted by the USPEX algorithm is validated using two machine learning potentials with fundamentally different architectures: the graph-based DPA-3 and DeepMD-se\_a, which relies on local atomic descriptors. If the discrepancies in the predicted values of enthalpy, forces, or energy exceed predefined thresholds, the given configuration is recalculated using density functional theory (DFT) and added to the training database. After the database has been updated, both potentials (DPA-3 and DeepMD-se\_a) are retrained, and the SUCCESS algorithm is restarted using the refined models for subsequent exploration.
}
    \label{fig:active_learn}
\end{figure*}

First-principles calculations of the selected structures used to construct the training database were performed within the framework of density functional theory (DFT) using the VASP package.

The exchange–correlation functional was treated within the generalized gradient approximation (GGA) using the PBE parametrization. The electronic structure was described using projector augmented-wave (PAW) pseudopotentials: Ca\_sv with 10 valence electrons ($3s^2 3p^6 4s^2$), Fe\_sv with 16 valence electrons ($3s^2 3p^6 3d^7 4s^1$), and Ni\_pv with 16 valence electrons ($3p^6 3d^9 4s^1$).

In plane-wave DFT calculations, the accuracy is largely controlled by the kinetic energy cutoff $E_{\mathrm{cut}}$, which determines the completeness of the basis set. Under high compression, the electronic density between closely spaced atoms develops high-gradient oscillations, requiring an increased basis-set resolution for accurate description.

To account for this effect, the cutoff energy was scaled with external pressure $P$. Since the characteristic wave vector $G$ required to resolve density oscillations scales inversely with the interatomic distance, and the kinetic energy scales as $|G|^2$, a linear approximation was adopted:
\[
E_{\mathrm{cut}}(P) = E_{\mathrm{cut}}(0) + \kappa P.
\]
The coefficient $\kappa$ was calibrated to ensure convergence of total energies and stress tensors across the entire pressure range from 0 to 500~GPa. For the Ca–Fe–Ni system, convergence tests yielded $E_{\mathrm{cut}}(0) = 500$~eV and $E_{\mathrm{cut}}(500~\mathrm{GPa}) = 1000$~eV, corresponding to $\kappa = 1$~eV/GPa.

This scaling strategy preserves high numerical accuracy along the full compression path without introducing unnecessary computational overhead at lower pressures.

All calculations were spin-polarized with a ferromagnetic initial magnetic configuration. Numerical precision was controlled using the parameters PREC = ACCURATE, an electronic self-consistency convergence criterion $E_{\mathrm{diff}} = 10^{-6}$~eV, and LASPH = TRUE to include non-spherical contributions from gradient corrections inside the PAW spheres.

Brillouin-zone sampling was generated automatically using the K-Point Grid Server with MINDISTANCE = 35~\AA\ (accounting for magnetic symmetry), corresponding to approximately 3000 $k$-points per reciprocal atom.

These settings ensure accurate Brillouin-zone integration and enable enthalpy determination with an accuracy of approximately 3~meV/atom over a broad pressure range.

\subsection{Fine-Tuning of the DPA-3 Foundation Model}

Based on the DFT-recalculated structures, a training database comprising approximately 6800 configurations was constructed. Training a new MLIP from scratch for a ternary system spanning an ultra-wide pressure range, particularly with calcium exhibiting anomalous compressibility, would require substantially larger datasets. However, the available dataset is sufficient for fine-tuning a pretrained FM.

For this purpose, we selected the DeePMD framework with the DPA-3 architecture, which provides a well-established, efficient, and robust fine-tuning procedure, enabling accurate adaptation of a generalized foundation model to the specific characteristics of the present system.

The DPA-3 model is built upon a multilayer graph neural network based on the Linear Graph Series (LiGS) architecture, specifically designed for large-scale FM. A key feature of this architecture is its ability to efficiently learn from heterogeneous datasets without requiring structural modification of the network. This property enables the development of transferable universal interatomic potentials applicable across a broad range of chemical systems and thermodynamic conditions~\cite{DPA3_bohrium}.

 In the present work, a pretrained DPA-3 FM was employed, consisting of two principal components (see Fig.~\ref{fig:dpa3_scheme}):
(i) a universal graph-based descriptor network pretrained on large materials databases, including Materials Project, and
(ii) a task-specific fitting network that maps descriptor outputs to total energies.

For fine-tuning, we adopted the fitting network configuration optimized for Materials Project–style DFT data (\verb|MP_traj_v024_alldata_mixu|), ensuring methodological consistency with our DFT calculations. The presence of numerous Ca–Fe–Ni-related compounds in the Materials Project database further improves transferability, as the pretrained model already encodes chemically relevant interactions, while fine-tuning adapts the energy landscape to high-pressure regimes.

Training was performed for 200,000 iterations using the standard DEEPMD loss-weight scheduling parameters. To improve virial accuracy — which is critical for reliable enthalpy evaluation at high pressure — the weighting coefficients associated with the virial term in the loss function were increased by more than an order of magnitude. This modification reduced errors in the $PV$ contribution by approximately 1.5×, thereby improving phase ranking stability under compression.

After fine-tuning, the model achieved a mean absolute error (MAE) in energy of $1.13 \times 10^{-2}$~eV/atom on the test set. Detailed accuracy metrics for energies, forces, and virials at different pressures are summarized in Table~\ref{tab:dpa3_ac}. The Table reveals a discrepancy between MAE and RMSE values, which arises primarily from the inclusion of configurations generated at ultra-high pressures (above 500~GPa). Although such structures represent a small fraction of the dataset and DFT accuracy decreases in this regime, their inclusion is methodologically important. They ensure physically correct short-range repulsion and prevent artificial atomic collapse during evolutionary searches. When these extreme configurations are excluded from the test set, MAE and RMSE become comparable, confirming that the dominant contribution to the error originates from highly compressed states (see Table~\ref{tab:dpa3_ac}).

\begin{figure*}[t]
    \centering
    \includegraphics[width=0.76\textwidth]{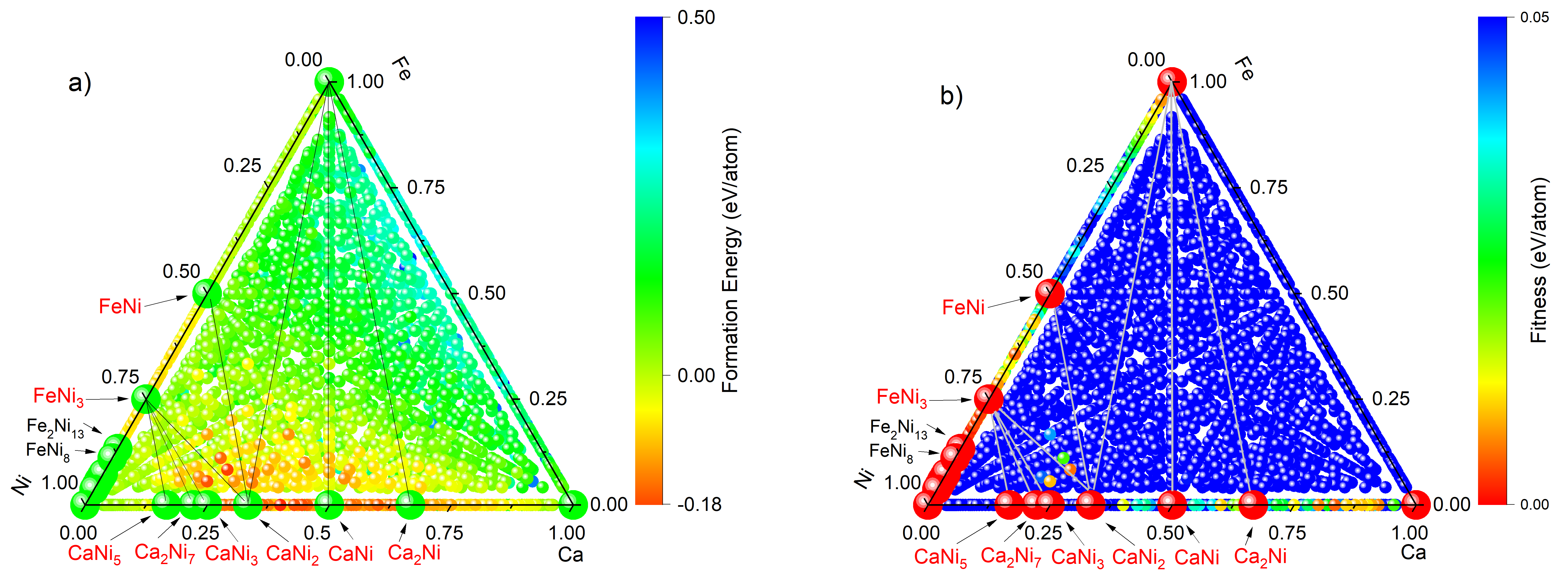} \\
    \caption{ Distribution of stable and metastable phases in the Ca–Fe–Ni system at ambient pressure ($P = 0$~GPa) in composition–energy coordinates. Configurations located near the convex hull were recalculated using VASP, while the remaining data were obtained using the foundation model (FM). Stable binary compositions are highlighted in red, while selected known stable solid solutions are indicated in black. In panel (a), the color scale represents the formation energy. In panel (b), the color encodes the distance to the convex hull (denoted as Fitness), characterizing the degree of metastability. All states with $\mathrm{Fitness} > 0.05$~eV/atom are shown in blue.}
    \label{fig:CaFeNi_P=0}
\end{figure*}

\subsection{Cross-Architecture Active Learning Strategy}

\begin{figure*}[thb]
    \centering
    \includegraphics[width=0.76\textwidth]{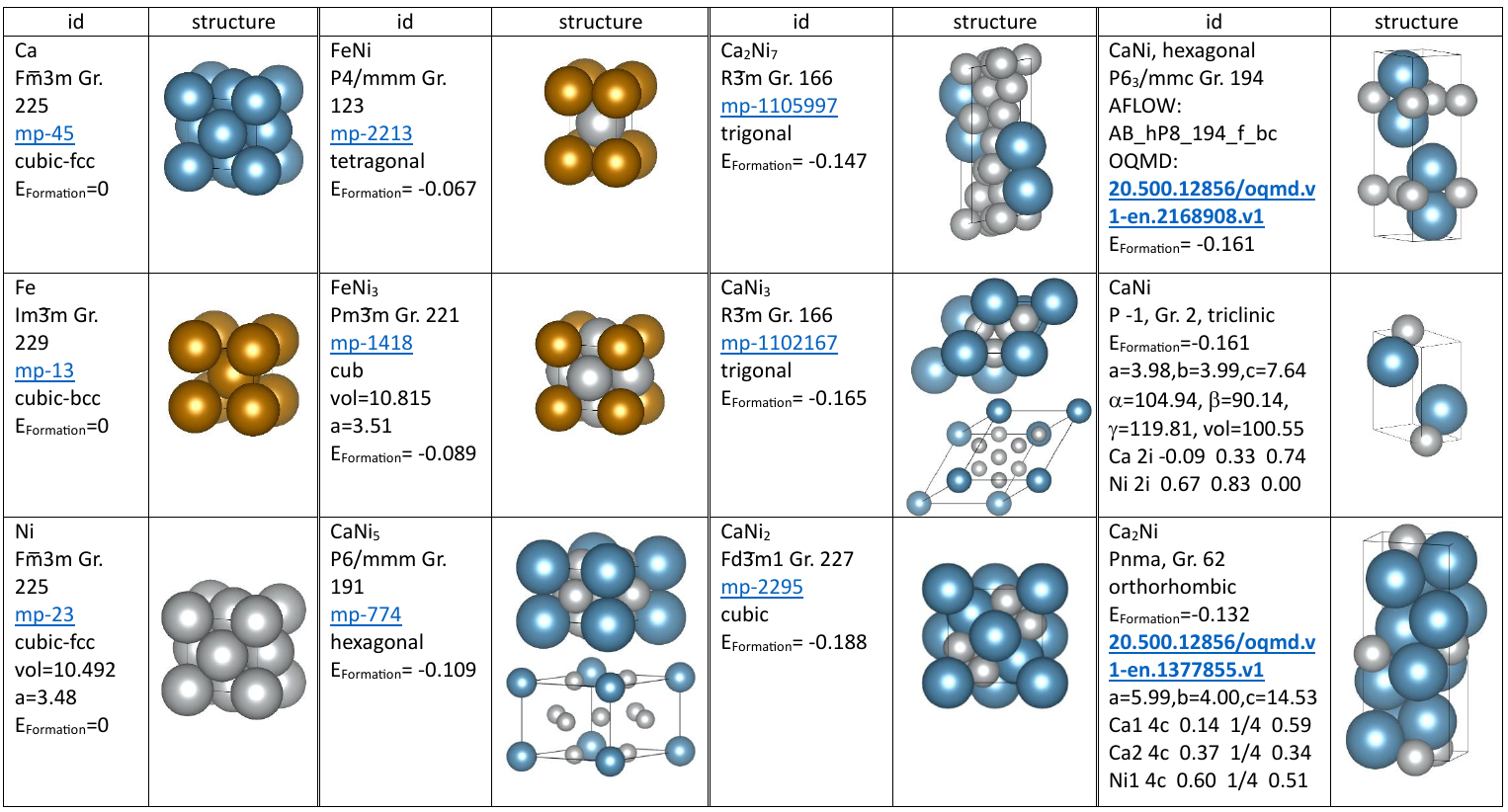} \\
    \caption{Crystal structures located on the convex hull of the Ca–Fe–Ni system at $P = 0$~GPa. Formation energies are given in eV/atom and lattice parameters in \AA. Structures reported in the Materials Project (mp-...), OQMD and AFLOW databases are labeled with their corresponding identifiers. The CaNi phase is not present in these databases; lattice parameters and atomic positions are provided.
}
    \label{fig:Convex_P=0}
\end{figure*}


\begin{figure*}[tbh]
    \centering
    \includegraphics[width=0.7\textwidth]{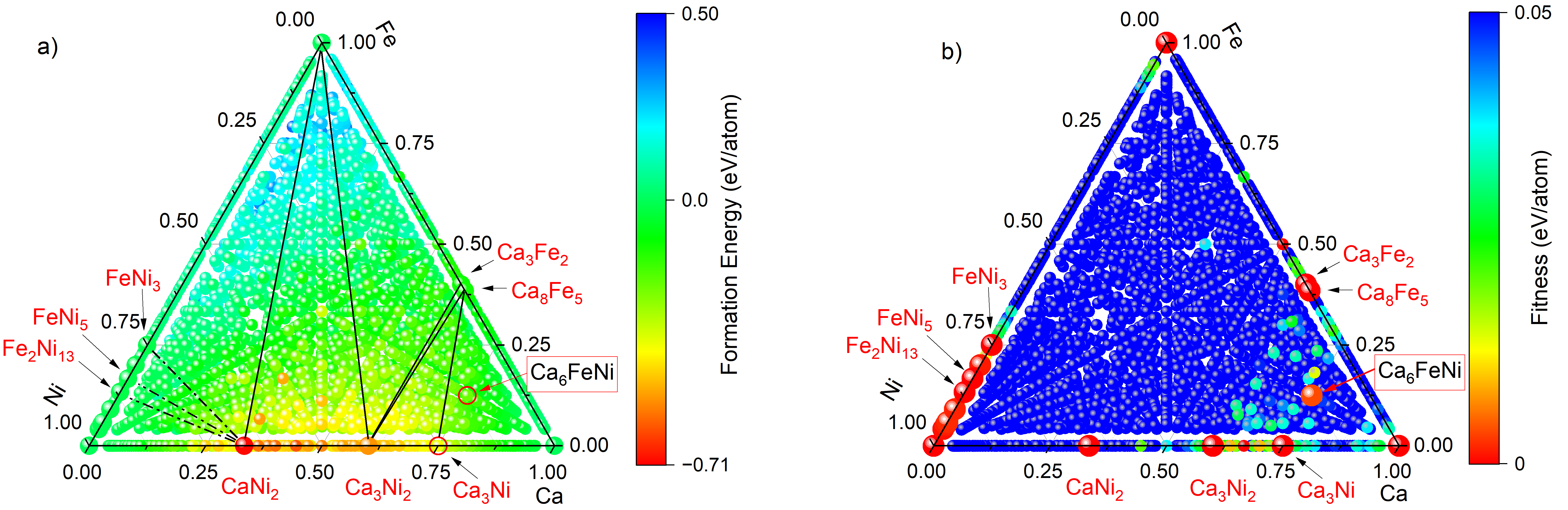} \\
\caption{Distribution of stable and metastable phases in the Ca–Fe–Ni system at $P = 50$~GPa. The color scale represents (a) formation energy and (b) Fitness, defined analogously to Fig.~\ref{fig:CaFeNi_P=0}.}
    \label{fig:CaFeNi_P=50}
\end{figure*}

\begin{figure*}[tbh]
    \centering
    \includegraphics[width=0.7\textwidth]{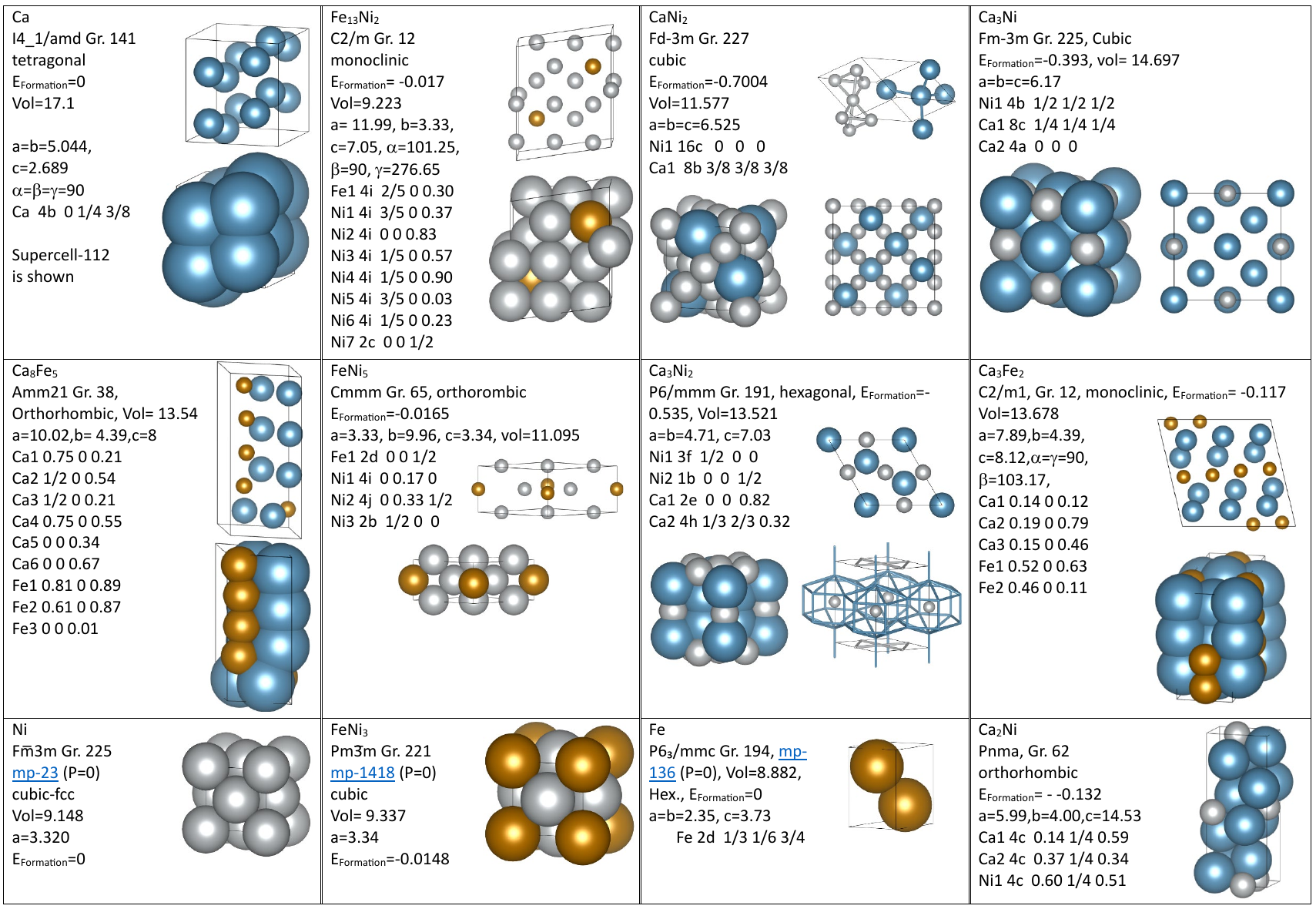} \\
\caption{Crystal structures located on the convex hull of the Ca–Fe–Ni system at $P = 50$~GPa (binary compounds are highlighted in Fig.~\ref{fig:CaFeNi_P=50}). Formation energies are given in eV/atom and lattice parameters in \AA. (In the table now: Fe13Ni2, should be: Fe2Ni13!!!!) (There are no designations for Wyckoff positions for Ca3Fe2!!!!!!!)}
    \label{fig:Convex_P=50}
\end{figure*}

Active learning is required to prevent accuracy degradation of the fine-tuned DPA-3 foundation model (FM) when encountering configurations whose topology or bonding environment differs substantially from those represented in the training dataset.

The conventional active-learning strategy for neural-network interatomic potentials relies on an ensemble of independently initialized models. The variance of predicted energies, forces, or virials is used as an uncertainty indicator: configurations with variance exceeding a predefined threshold are recalculated using DFT and incorporated into the training set. This approach is implemented, for example, in the DP-GEN framework.

However, in the present work, the model is obtained via fine-tuning of a pretrained FM rather than training from random initialization. Since all fine-tuned models inherit nearly identical pretrained weights, constructing a statistically diverse ensemble is not feasible. Even when only the fitting network was randomly reinitialized prior to fine-tuning, the resulting ensemble variance in predicted energies remained below 1~meV/atom, rendering variance-based uncertainty metrics ineffective.

To overcome this limitation, we introduce a conceptually different active-learning criterion based on cross-architecture disagreement.

In addition to the graph-based DPA-3 FM, a second interatomic potential was trained on the same DFT dataset using the DeepMD-se\_a\_ebd\_v2 architecture. In contrast to graph neural networks, this model relies on local atomic descriptors constructed from radial and angular information within a finite cutoff radius. Owing to its fundamentally different representation of atomic environments, it exhibits distinct extrapolation behavior outside the training manifold.

For every structure generated during the USPEX search — including those present in the training set — predictions of the DPA-3 and DeepMD-se\_a models were compared. Configurations for which the disagreement in predicted energy per atom, forces, or enthalpic contribution $PV$ exceeded predefined thresholds were selected for DFT recalculation (Fig.~\ref{fig:active_learn}). These structures were additionally subjected to small random perturbations prior to DFT evaluation to enrich the local sampling of the potential-energy surface.

This procedure defines a new uncertainty metric: instead of statistical variance within a single architectural family, uncertainty is quantified by systematic disagreement between two physically distinct model classes. The criterion is particularly suitable for fine-tuned foundation models, where ensemble diversity is inherently suppressed.

The active-learning loop converged within three iterations. In the first cycle, applying the threshold $| \Delta PV | > 0.04$~eV/atom identified several hundred configurations requiring DFT refinement. After three iterations, the number of high-disagreement structures ($| \Delta PV | \sim 0.05$~eV/atom) decreased to approximately ten. The total training dataset increased by roughly 500 configurations.

Following the initial active-learning stage, the refined DPA-3 FM was employed for evolutionary searches at progressively increasing pressures: $P = 0$, $50$, $100$, $200$, and $400$~GPa. At each pressure, newly generated structures were passed through the same cross-architecture filtering procedure using a stricter threshold of $| \Delta PV | > 0.03$~eV/atom. This iterative refinement ensures that configurations critical for phase ranking under compression are systematically incorporated into the training dataset.

Importantly, while global accuracy metrics (see Fig.~\ref{fig:MLP_acc} and Table~\ref{tab:dpa3_ac}) changed only marginally, the active-learning strategy substantially improved model robustness. In particular, it eliminated spurious low-energy configurations arising from extrapolation artifacts and ensured stable behavior under extreme compression.

The proposed cross-architecture active-learning criterion constitutes a key methodological contribution of this work and provides a general strategy for uncertainty control in fine-tuned foundation-model–assisted crystal structure prediction.

\section{Results}

\subsection{$P = 0$~GPa}

As an initial validation step, the Ca--Fe--Ni system was investigated at ambient pressure using the USPEX evolutionary algorithm coupled with three different foundation models: MACE, MatterSim, and the fine-tuned DPA-3 model.

All three approaches yielded consistent sets of low-energy structures in the vicinity of the convex hull. For final verification, structures with formation energies within 50~meV/atom above the convex hull were recalculated using DFT in VASP, and the convex hull was reconstructed based on the DFT energies.

The resulting phase diagram (Figs.~\ref{fig:CaFeNi_P=0}–\ref{fig:Convex_P=0}) is in agreement with available literature and database data. Stable compositions are indicated on the convex hull, while selected metastable phases are also shown for completeness. In particular, metastable states within 10~meV/atom of the hull correspond to Fe-rich solid solutions in Ni.

The stable elemental reference phases are correctly reproduced: Ca adopts the fcc structure under ambient conditions~\cite{Maksimov2005,Oganov2010}, Fe stabilizes in the bcc structure, and Ni in the fcc structure. No thermodynamically stable ternary compounds were identified at $P = 0$~GPa.

Binary compounds located on the convex hull include FeNi, FeNi$_3$, CaNi$_5$, Ca$_2$Ni$_7$, CaNi$_3$, and CaNi$_2$, all consistent with reported phases. The CaNi compound, present in the AFLOW database, is also confirmed to lie on the convex hull.

In addition to the known hexagonal CaNi phase reported in AFLOW, we identified a new triclinic CaNi structure with nearly identical formation energy (difference below 1~meV/atom). The difference in energy (per atom) between these phases is very small: $E_{\rm hex} = -3.87143$~eV and $E_{\rm tri} = -3.87224$~eV. So the triclinic structure is more stable. It also exhibits a more isotropic unit cell, with comparable lattice parameters $a$, $b$, and $c$, whereas the hexagonal phase is characterized by a highly elongated $c$ axis. The triclinic phase is also denser, with an atomic volume of 25.127~\AA$^3$/atom compared to 25.189~\AA$^3$/atom for the AFLOW phase (see Fig.~\ref{fig:Convex_P=0}).

\begin{figure}[htb]
  \centering
  \includegraphics[width=0.6\columnwidth]{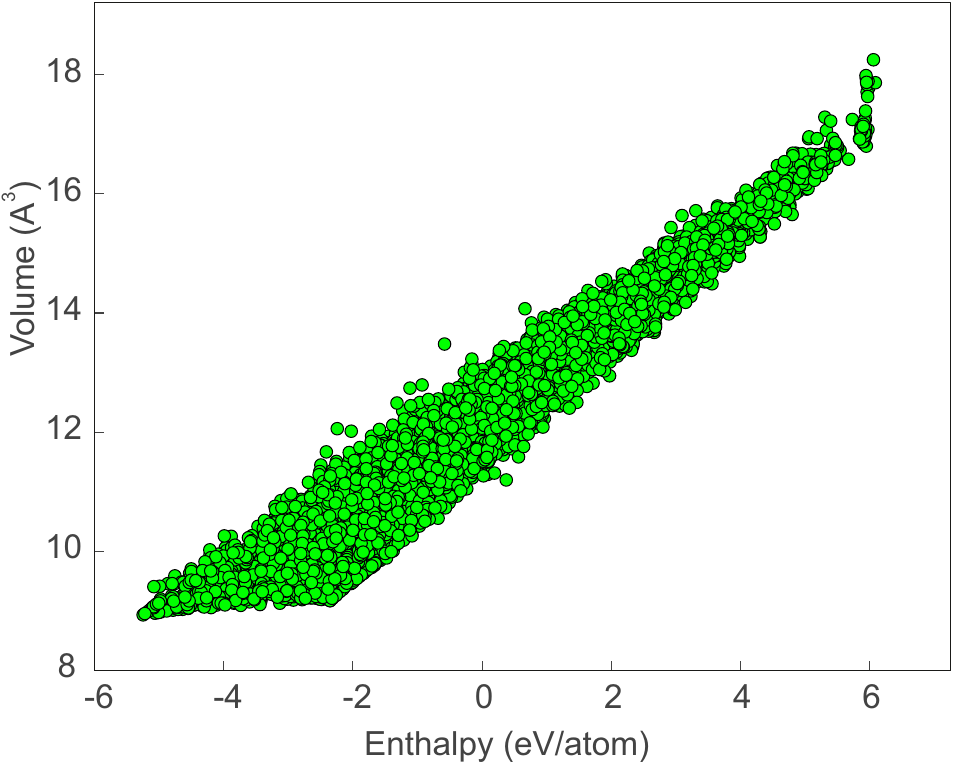}
  \caption{Atomic volume and enthalpy per atom for structures identified by the USPEX algorithm at $P = 50$~GPa.}\label{fig:Enthalpy_V_P=50}
\end{figure}

\begin{figure*}[tbh]
    \centering
    \includegraphics[width=0.7\textwidth]{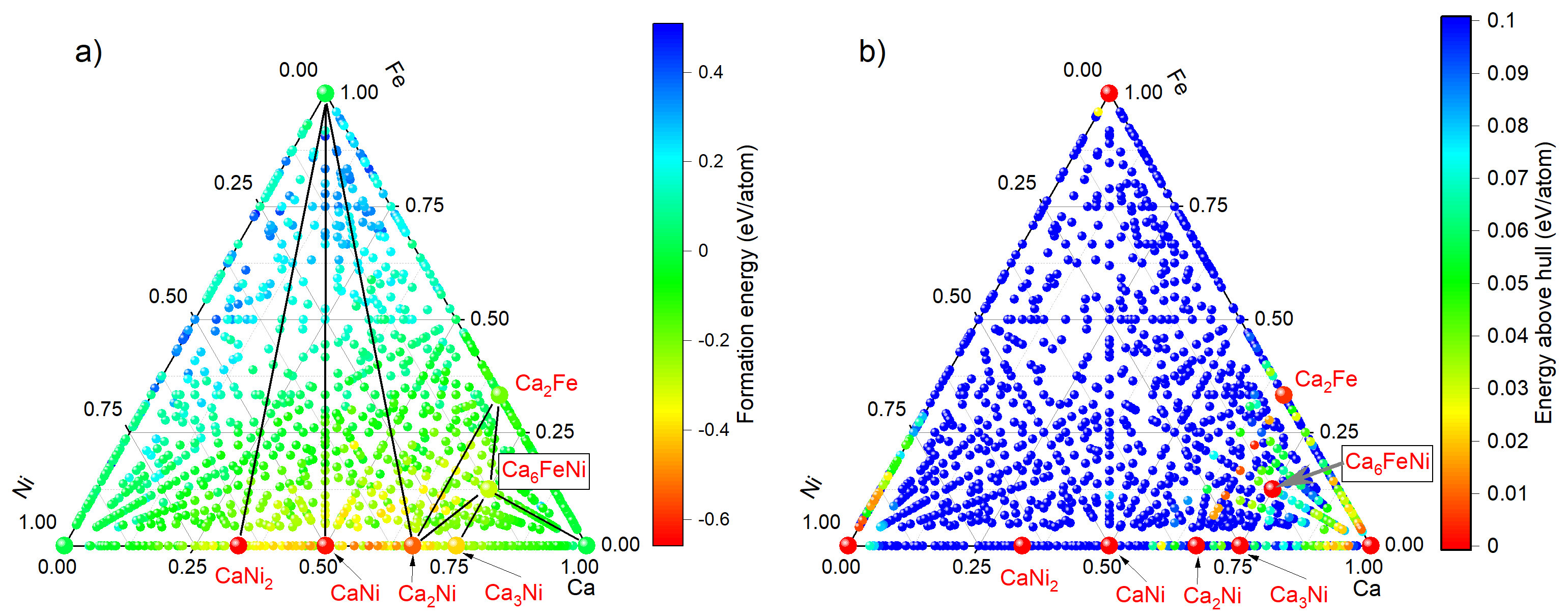} \\
\caption{Distribution of stable and metastable phases in the Ca–Fe–Ni system at $P = 100$~GPa. The color scale represents (a) formation energy and (b) Fitness, defined analogously to Fig.~\ref{fig:CaFeNi_P=0}.}
    \label{fig:CaFeNi_P=100}
\end{figure*}

\begin{figure*}[tbh]
    \centering
    \includegraphics[width=0.7\textwidth]{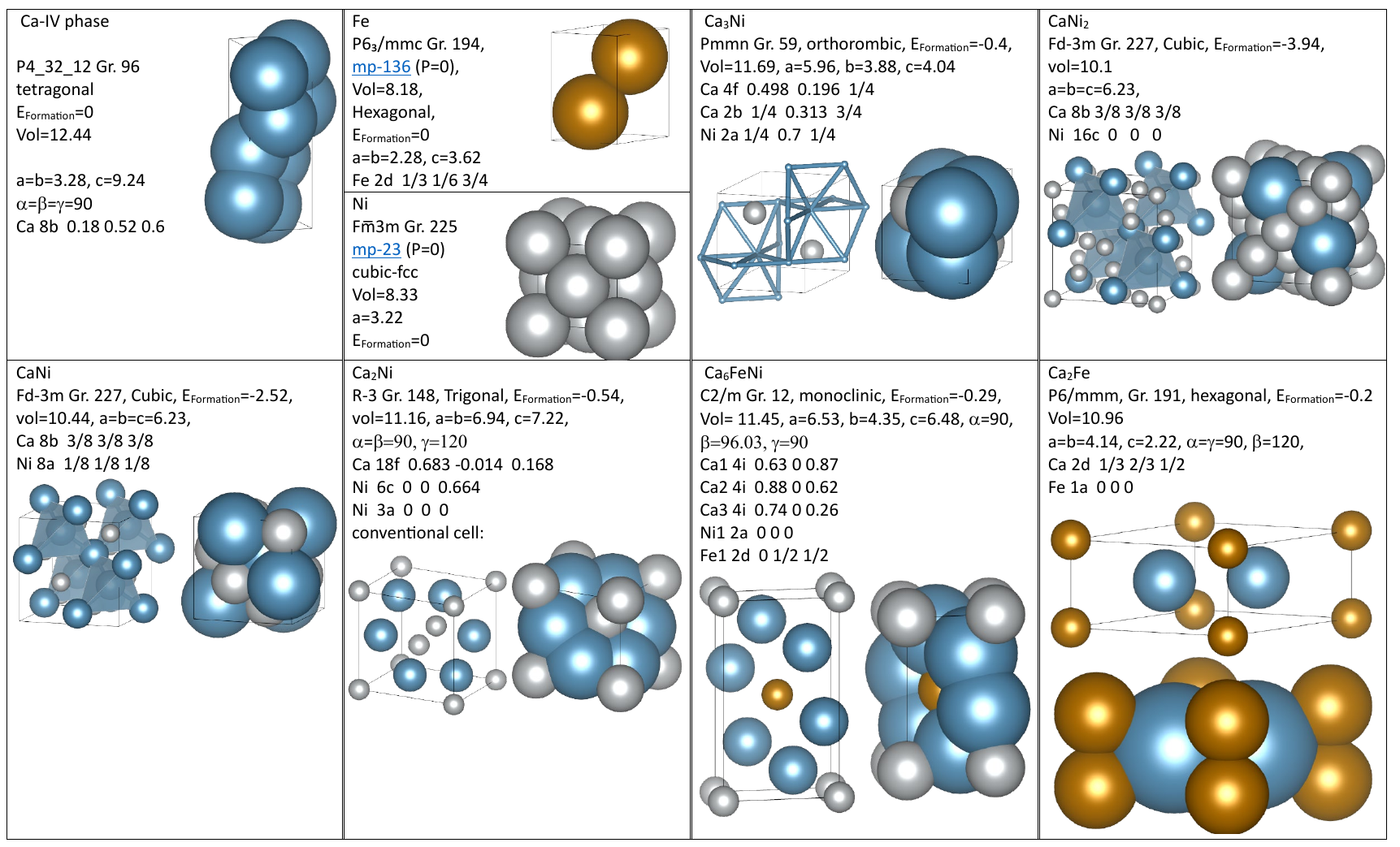} \\
\caption{Crystal structures located on the convex hull of the Ca–Fe–Ni system at $P = 100$~GPa (binary compounds are highlighted in Fig.~\ref{fig:CaFeNi_P=100}). Formation energies are given in eV/atom and lattice parameters in \AA.}
    \label{fig:Convex_P=100}
\end{figure*}
\subsection{$P = 50$~GPa}

At elevated pressures, the USPEX algorithm operates by comparing enthalpies rather than total energies, selecting thermodynamically favorable structures accordingly. A standard validation of the evolutionary search consists in analyzing the dependence of enthalpy on atomic volume for all generated structures.

Figure~\ref{fig:Enthalpy_V_P=50} shows the enthalpy–volume distribution for structures identified by USPEX, where each point corresponds to a distinct configuration. The presence of a well-defined, smooth dependence (approximately linear within certain regions) indicates the consistency of the evolutionary search, as well as the correctness of the implemented computational workflow. Such behavior is expected for systems approaching thermodynamic equilibrium and suggests that the configuration space has been sampled in a sufficiently systematic manner.

At $P = 50$~GPa and zero temperature, iron stabilizes in the hexagonal close-packed (hcp) structure, while nickel retains its face-centered cubic (fcc) phase, consistent with its behavior at ambient pressure. Calcium adopts the $I4_1/amd$ (space group 141) structure, which can be interpreted as a distorted derivative of the simple cubic phase. This phase is known to be thermodynamically stable in the pressure range of approximately 33–71~GPa and differs from the ideal simple cubic structure~\cite{Oganov2010}. The present DFT calculations at $P = 50$~GPa are in full agreement with these observations.

At ambient pressure, iron and calcium do not form stable compounds. Interestingly, pressure stabilizes Ca--Fe intermetallics with a noticeable iron content $\stackrel{\sim}{=}$ 40\%: Ca$_3$Fe$_2$ and Ca$_8$Fe$_5$. The structure of Ca$_3$Fe$_2$ is nontrivial, despite the phase having low symmetry: iron and calcium form ``chains,'' with iron wedging into the voids between the chains of calcium atoms. The Ca$_8$Fe$_5$ phase exhibits a similar arrangement, except that its calcium chains are not ``straight'' but rather deformed, see Fig.~\ref{fig:Convex_P=50}.

In Figure~\ref{fig:CaFeNi_P=50}, the ternary phase Ca$_6$FeNi is highlighted. It remains metastable at a pressure of 50~GPa but lies very close to the convex hull, with a distance of 0.0038~eV/atom. At 100~GPa, this phase becomes stable.

\subsection{$P = 100$~GPa}

The most exciting results are obtained at a pressure of $P=100$~GPa. In this case, the ternary phase Ca$_6$FeNi turns out to be stable, as shown in Fig.~\ref{fig:CaFeNi_P=100}. The structure of this phase is quite unusual. In the unit cell, calcium atoms form an almost regular hexagon with nearest-neighbor distances of 2.44~\AA. The center of this hexagon provides sufficient space to accommodate a single iron atom. This entire arrangement is framed by nickel atoms, which form the vertices of a monoclinic cell (slightly distorted tetragonal cell).

Another interesting point is the formation of a new stable binary iron–calcium phase, Ca$_2$Fe. This phase is highly symmetric (hexagonal).

The results at $P = 100$~GPa were obtained faster than those at $P = 0$ and $50$~GPa. This is because we used, as seeds, the 2000 structures closest to the convex hull obtained from the calculations at $P = 0$ and $50$~GPa. Consequently, the USPEX algorithm converged much more rapidly. The convergence criterion was the stabilization of the convex hull and the absence of new stable structures appearing within $0.05$~eV/atom of the convex hull for 15 consecutive generations.

\section{Conclusions}




 \bibliographystyle{elsarticle-num}
 \bibliography{bib_CaFeNi}





\end{document}